\documentclass[aps,pra,twocolumn,floatfix,showpacs]{revtex4}

\usepackage{graphics}
\usepackage{bm}

\def\PRA{{Phys.~Rev.~A} }

\def\JPB{{J.~Phys.~B} }
\def\PRL{{Phys.~Rev.~Lett.} }
\def\RMP{{Rev.~Mod.~Phys.} }

\newcommand{\myscalebox}[1]{\scalebox{0.45}[0.45]{#1}}

\newcommand{\be}{\begin{equation}}
\newcommand{\bea}{\begin{eqnarray}}
\newcommand{\eea}{\end{eqnarray}}
\newcommand{\ee}{\end{equation}}

\begin{document}

\title{Theoretical analysis of dynamic chemical imaging with lasers
using high-order harmonic generation }

\author{Van-Hoang Le,$^{1,2}$ Anh-Thu Le,$^1$ Rui-Hua Xie,$^1$ and C.~D. Lin$^1$}

\affiliation{$^1$J. R. Macdonald Laboratory, Department of Physics,
Cardwell Hall, Kansas State University, Manhattan, KS 66506, USA\\
$^2$Department of Physics, University of Pedagogy, 280 An Duong
Vuong, Ward 5, Ho Chi Minh City, Vietnam\\ }

\date{\today}

\begin{abstract}

We report theoretical investigations of the tomographic procedure
suggested by Itatani {\it et al.} [Nature, {\bf 432} 867 (2004)]
for reconstructing highest occupied molecular orbitals (HOMO)
using high-order harmonic generation (HHG).  Using the limited
range of harmonics from the plateau region, we found that under
the most favorable assumptions, it is still very difficult to
obtain accurate HOMO wavefunction, but the symmetry of the HOMO
and the internuclear separation between the atoms can be
accurately extracted, especially when lasers of longer wavelengths
are used to generate the HHG. We also considered the possible
removal or relaxation of the approximations used in the
tomographic method in actual applications. We suggest that for
chemical imaging, in the future it is better to use an iterative
method to locate the positions of atoms in the molecule such that
the resulting HHG best fits the macroscopic HHG data, rather than
by the tomographic method.

\end{abstract}

\pacs{42.65.Ky, 33.80.Rv}

\maketitle

\section{Introduction}

  In the microscopic world, chemical reactions and biological transformations
  occur on the time scale of the order of picoseconds or less. Understanding the
   intermediate steps of these reactions or transformations has been
   the dream of physical and chemical scientists \cite{zewail,zewail01}. While X-ray
   and electron diffraction have served well to image the structure of large
   molecules, current technology limits such tools to  time resolution to
   tens of picosecond to sub-picoseconds \cite{zewail06,suzuki06}, making
   them less suitable for time-dependent studies. On the other hand, infrared lasers
   of durations of tens to sub-ten femtoseconds are becoming widely available today, thus
   it is natural to ask whether femtosecond lasers can be used for dynamic
   chemical imaging of molecules.
   Since the wavelength of infrared lasers is much larger than the
   interatomic spacings, these lasers cannot be used for diffraction measurement
   directly. Nevertheless, recent experiments \cite{cocke} have shown that the
   shape of the highest occupied molecular orbitals (HOMO) can be revealed by
   measuring the alignment-dependence of the ionization rates by sub-ten
   femtosecond lasers. Similarly, limited experimental evidences and
   theoretical calculations so far have also shown that the high-order harmonic
   generation (HHG) can reveal the structure of molecules
   \cite{zhou,lein,baker}, including even two-center interference
   \cite{lein02,kanai,marangos}. Self-diffraction from the
    rescattering electron following laser ionization has also been
    suggested for imaging molecules \cite{spanner04,yurchenko,hu}. Despite many such
    discussions in the literature in the past decade, few real
    demonstrations have been achieved so far.

   In a recent paper,
   Itatani {\it et al.} \cite{itatani-nature} reported that they
   have  reconstructed the highest occupied molecular orbital
   of N$_2$ molecules accurately using the tomographic procedure
   from the measured HHG spectra.
   Using a weak laser beam first to partially
   align  N$_2$ molecules, a more intense short probe pulse (30 fs)
   was then used to generate high-order harmonics.
The polarization direction of the probe pulse was varied  from
parallel to perpendicular direction with respect to the pump pulse.
The retrieved HOMO wavefunction has been shown to be in surprisingly
good agreement with the HOMO calculated from the quantum chemistry
code. In a more recent paper, taking that the HHG are generated from
a many-electron N$_2$ molecule, it was further shown \cite{patch06}
that the reconstructed wave-function records the image of the Dyson
orbital plus exchange contributions from the inner shells.

 The success of Itatani {\it et al.} has generated a lot of interest.
 If it is generally applicable, it would offer
 the means for  time-resolved dynamic chemical imaging
 at the resolution of ten to sub-ten femtoseconds. However, a number of
skepticism have been raised recently \cite{schwarz,greene}.
Rigorously speaking, in quantum mechanics, the ground state
wavefunction is not measurable alone since any measurement would
invoke final states. The tomographic procedure used by Itatani
{\it et al.} relies on the approximation that the continuum states
of the molecule in the recombination process be represented by
plane waves, an approximation that is well known to be invalid in
photochemistry, unless the continuum electron's energy is in the
keV range. This is not the case for HHG in the plateau region
where the photons are from  a few tens to about 100 eV.

The starting point of Itatani {\it et al.} is the three-step model
of high-order harmonic generation
\cite{kulander,corkum,lewenstein}. The harmonic yield is assumed
to be given by the product of three terms, one is the initial
tunneling ionization rate, the next is the propagation of the
ionized electron, and the third  is the recombination of the
electron with the ion to emit high-order harmonics.   By assuming
that the first two steps are nearly independent of whether the
target is molecular or atomic so long they have nearly identical
ionization potentials, Itatani {\it et al.} were able to extract
the recombination dipole matrix element for individual N$_2$
molecules from the measured HHG in the plateau region, as well as
their variation with angles. These dipole matrix elements are then
used to obtain the molecular wavefunction using the tomographic
imaging method.

  Two important approximations are used in   this tomographic procedure.
  First, the dipole moment of individual molecules in the direction
  parallel and perpendicular to laser's polarization can be obtained
  from the measured macroscopic HHG. For this, they assumed that the
  harmonics in the direction parallel to the laser polarization
  satisfies complete phase matching condition. Experimentally this is done
  by placing the gas jet after the focus of the laser, thus eliminating the
  contribution to HHG from the long trajectories \cite{antoine,balcou}.
  They did not measure HHG in the direction perpendicular to the laser's
  polarization, but the dipole moment in the perpendicular direction was
  assumed. In practice, if the molecules are fully aligned, i.e.,
  all pointing in the same direction in space, then the measurement of macroscopic
  HHG in the perpendicular direction would allow the determination of the dipole
  moment in the perpendicular direction.   For
  molecules that are only partially aligned, such as those aligned by a
  pump laser, it is not straightforward to extract the perpendicular
  dipole moment of each molecule from the measured macroscopic
  HHG.

  The second important approximation in Itatani {\it et al.} is that the continuum
  wavefunctions in the recombination dipole matrix elements can be
  approximated by plane waves. This approximation is at the heart of the
  tomographic procedure. Implicit to this also is that the dipole moment is
  available over the complete spectral range of the harmonics. This is not so since the
useful harmonics are  usually limited to the plateau region only.

In this paper we examined the assumptions used in the tomographic
procedure. We used the Lewenstein model \cite{lewenstein} to
calculate the HHG's for fixed molecules with different alignment
angles with respect to the laser polarization. In this model, the
dipole matrix element is calculated using plane waves for the
continuum states. We thus take the calculated HHG from the
Lewenstein model to be the ``experimental" data, and follow the
procedure of Itatani {\it et al.} to extract the ``experimental"
dipole matrix elements. If the three-step model is exact, the
``experimental" dipole matrix element thus obtained should be
identical to the ``exact" theoretical ones which are calculated
using plane waves for the continuum states and the HOMO for the
initial state. In reality, we found that the ``experimental"
dipole matrix element agrees with the theoretical one only in a
limited spectral range -- in the plateau region. With such a
limited range, we found that the  HOMO wavefunctions derived from
the tomographic procedure to be approximate. Even though the
accuracy can be improved by using HHG generated with lasers of
longer wavelength to extend the plateau region, we were unable to
obtain accurate HOMO wavefunction.

In view of the failure in obtaining accurate HOMO wavefunctions, we
next turn to the possibility of using HHG to extract some
information on the structure of a molecule.  We note that in
  chemical imaging it is not essential to
reconstruct the wavefunction of a molecule. A great deal is
understood if the positions of the atoms in the molecule  can be
determined at the intermediate time steps. We thus ask what
partial information of a molecule can be revealed by the
tomographic procedure under the most favorable conditions assumed
in our model. Using lasers with wavelength of 1200 nm to generate
plateau harmonics from N$_2$ and O$_2$
 molecules at different internuclear distances, we showed that the
 internuclear distances obtained from the retrieved HOMO
 wavefunction is good to better than a few percents. Furthermore,
 the geometry and the symmetry of the HOMO, including
 the nodal surfaces, can also be retrieved. Such precision is
more than adequate for dynamic chemical
 imaging and warrants further test for complex molecules where the
 positions of some atoms undergo large change, in processes such
 as isomerization or dissociation, and where the time scale of
 transformation is of the order of  tens of femtoseconds.

  The rest of this paper is arranged as follows. In
Section II we describe the theoretical methods and assumptions
used for the present tomographic imaging of molecules using
high-order harmonic generation of aligned molecules by lasers.
Section III presents the results of extracting HOMO wavefunctions
and the internuclear separations. In Section IV we revisit the
assumptions  and address the intrinsic limitations of the
tomographic imaging method. We showed that these limitations are
not easily removed in real applications, making it difficult to
apply the tomographic method for chemical imaging in general. In
Section V, we suggest alternative methods for  imaging molecules
with HHG, based on the iterative procedure. Assuming that the HHG
for a single polyatomic molecule can be calculated accurately, the
propagation effect and the effect of partial alignment of the
molecules can be  taken into account by the standard methods. Thus
  macroscopic HHG can be calculated for each configuration of
  atoms in the   molecule which can then be compared to the
experimentally measured HHG. Since HHG can be generated for
different alignment angles of the molecules using lasers of
different wavelengths, an optimum atomic configurations can be
derived by the set that best fits the experimental HHG. We note
that in typical chemical reactions, only a few atoms change
positions significantly. Thus the number of parameters that have
to be determined is small. We then give examples of the
isomerization of acetylene  C$_2$H$_2$ and of hydrogen cyanide HCN
to show the alignment dependence of the HHG of these molecules
under isomerization. In the last Section we summarize our
conclusion on the tomographic procedure and discuss future plan
for studying dynamic chemical   imaging with lasers. Atomic units
$m=e=\hbar=1$ are used throughout the paper, unless otherwise
indicated.

\section{Theoretical Methods}

\subsection{The generation of HHG using Lewenstein model}
In this paper, in order to generate the HHG spectra from molecules
with fixed alignments, we use the Lewenstein model, as extended by
Zhou {\it et al.} \cite{zhou}. Without loss of generality, we
assume that the molecules are aligned along $x$-axis in a laser
field $E(t)$, linearly polarized on the $x$-$y$ plane with an
angle $\theta$ with respect to the molecular axis. The parallel
component of the induced dipole moment can be written in the form
\begin{eqnarray}
D_{\parallel}(t)& = &
i\int_0^{\infty}d\tau\left(\frac{\pi}{\epsilon+i\tau/2}
\right)^{3/2}[\cos\theta d^*_x(t)+\sin\theta d^*_y(t)]\nonumber\\
 & &
\times[\cos\theta d_x(t-\tau)+\sin\theta d_y(t-\tau)]E(t-\tau)\nonumber\\
& & \times\exp[-iS_{st}(t,\tau)]a^*(t)a(t-\tau)+c.c. \label{model}
\end{eqnarray}
where ${\bm d}(t)\equiv{\bm d} [{\bm p}_{st}(t,\tau)-{\bm A}(t)]$,
${\bm d}(t-\tau)\equiv{\bm d} [{\bm p}_{st}(t,\tau)-{\bm
A}(t-\tau)]$ are the transition dipole moments between the ground
state and the continuum state, and ${\bm p}_{st}(t,\tau) =
\int_{t-\tau}^{t} \bm{A}(t') dt'/\tau$ is the canonical momentum
at the stationary points, with ${\bm A}$ the vector potential. The
perpendicular component $D_{\perp}(t)$ is given by a similar
formula with $[\cos\theta d^*_x(t)+\sin\theta d^*_y(t)]$ replaced
by $[\sin\theta d^*_x(t)-\cos\theta d^*_y(t)]$ in Eq.~(1).

  The quasiclassical action at the stationary points
for the electron propagating in the laser field is
\begin{equation}
S_{st}(t,\tau) = \int_{t-\tau}^{t} \left( \frac{
[\bm{p}_{st}(t,\tau)-\bm{A}(t')]^2}{2}+I_p\right) dt', \label{action}
\end{equation}
where $I_p$ is the ionization potential of the molecule. In
Eq.~(\ref{model}), $a(t)$ is introduced to account for the ground
state depletion.

The ground state electronic wavefunctions were obtained from the
GAMESS code \cite{gamess}. Within the Single Active Electron (SAE)
approximation, we take the HOMO for the ``ground state", the
transition dipole ${\bm d}({\bm k})$ is given as $\langle{\bm
k}|{\bm r}|\Psi_0({\bm r})\rangle$. The continuum state is
approximated here by a plane-wave $|{\bm k}\rangle$. This model
has been shown to be able to interpret the experimental results so
far for N$_2$, O$_2$ \cite{itatani,zhou} and CO$_2$
\cite{kanai,vozzi,atle06}. In this paper we neglect the effects of
depletion, i.e., we assume, $a(t)=a^*(t)=1$. Laser peak intensity
of $2\times 10^{14}$ W/cm$^2$ and duration of $30$ fs (FWHM) are
used throughout this paper.

\subsection{The procedure of tomographic imaging}
In the tomographic procedure, suggested by Itatani {\it et al.}
\cite{itatani-nature}, the intensity of the emitted HHG is
approximated by
\begin{equation}
S(\omega,\theta)=N^2(\theta)\omega^4 |a[k(\omega)]{\bm
d}(\omega,\theta)|^2. \label{intensity}
\end{equation}
Here ${\bm d}(\omega,\theta)=\langle\Psi_0({\bm r};\theta)|{\bm
r}|\exp[ik(\omega)x]\rangle$ is the transition dipole between the
HOMO and the continuum state, approximated by a plane-wave with
$k({\omega})=(2\omega)^{1/2}$; $a[k(\omega)]$ is the amplitude of
the continuum plane wave of returned electrons; and $N(\theta)$ is
the number of ions produced. $\Psi_0({\bm r};\theta)$ represents the
HOMO, rotated by angle $\theta$ (the Euler angle between molecular
axis and the laser polarization direction). Note that in
Eq.~(\ref{intensity}) perfect phase-matching was assumed in the
experimentally measured spectra. This equation can be understood as
a further approximation to the three-step model \cite{rmp-review}.
In the case of strong-field tunneling ionization, it can be assumed
that the ionized electron wave packet depends weakly on the
molecular structure, but rather on the ionization potential only.
Thus, one can get rid of $a[k(\omega)]$ by measuring the HHG from
the reference atoms with the same $I_{p}$ (for N$_2$ the reference
atom is Ar, for O$_2$ it is Xe).

From the measured HHG and ionization data one can extract the
transition dipole ${\bm d}(k,\theta)$ by using formula
\begin{equation}
|{\bm d}(k,\theta)|=N(\theta)^{-1}|{\bm d}_{\it
ref}(k)|\sqrt{S(\omega,\theta)/S_{\it ref}(\omega)}
\label{dipole1}
\end{equation}
with ${\bm d}_{\it ref}(k)$ being the transition dipole moment of
the reference atom. From Eq.~(\ref{dipole1}) we can extract only
the amplitude of the dipole moment. However, it is well-known
\cite{Hay} that the minimum of harmonic signal and a $\pi$ phase
jump occur at the same harmonic order. Following Itatani {\it et
al.}, we assume that the dipole changes sign when its absolute
value goes through a minimum close to $0$.

In this study we will start with HHG from ${\it individual}$
molecules which were calculated using the Lewenstein model. This
allows us to bypass the issues related to phase matching and other
macroscopic effects for the time being. Following Itatani {\it et
al.} we obtain the ${\bm d}(\omega,\theta)$ from
Eq.~(\ref{intensity}) except replacing $N^2(\theta)$ in the equation
by $N(\theta)$ (since phase matching is not used for a single
molecule). The ionization rates $N(\theta)$ are calculated using the
molecular tunneling theory (MO-ADK) \cite{mo-adk}.

One major difference from Itatani {\it et al.} is that in our
calculations we use the dispersion relation
\begin{equation}
k=\sqrt{2(\omega-I_{\it p})} \label{dispersion}
\end{equation}
with $I_p$ being the ionization potential. The question about
including or not including $I_p$ in Eq.~(\ref{dispersion}) was
discussed in a number of papers (see, for example,
\cite{kamta,david06}); it is a question of how to relate the energy
of the re-colliding electron with the energy of the emitted photon.
For radiative recombination or photoionization,
Eq.~(\ref{dispersion}) is the correct relation with $k$ being the
momentum of the electron in the asymptotic region. In the Lewenstein
model the re-colliding electron does not see any potential from the
ion so $k^2$/2 is also the kinetic energy of the electron near the
nucleus. Indeed we found that the deduced dipole moment from HHG
calculated using Lewenstein model works well only when
Eq.~(\ref{dispersion}) was used. In Itatani {\it et al.}, the
relation $k=\sqrt{2\omega}$ was used, but their dipole moment was
extracted from the experiment.

From the dipole moment parallel and perpendicular to the
polarization direction, we first rotate  the two components to the
molecular frame. The
  wave function is then obtained  by using
the Fourier slice theorem \cite{kak}.  The basic formulae for
tomographic procedure
\begin{eqnarray}
x\Psi(x,y)&= \int_{0}^{\pi} d\theta \int_{0}^{+\infty} d\omega
e^{\imath k (x \cos\theta+y \sin\theta)} \nonumber \\  & \times
[\cos\theta d_{x}(\omega,\theta)+\sin\theta d_{y}(\omega,\theta)]
\label{tomo1}
\end{eqnarray}
\begin{eqnarray}
y\Psi(x,y)&= \int_{0}^{\pi} d\theta \int_{0}^{+\infty} d\omega
e^{\imath k (x \cos\theta+y \sin\theta)} \nonumber \\  & \times
[-\sin\theta d_{x}(\omega,\theta)+\cos\theta d_{y}(\omega,\theta)]
\label{tomo2}
\end{eqnarray}
can then be used to obtain the HOMO in the molecular frame. Note,
here and in the following for the wave-function integrated along
the third direction we use the notation
\begin{equation}
\Psi(x,y)\equiv \int_{-\infty}^{+\infty}\Psi_0(x,y,z)dz.
\label{integrate-wf}
\end{equation}
In principle, the two equations (\ref{tomo1}) and (\ref{tomo2})
are equivalent. In practice, however, incomplete dipole moment
data would give different results from equations (\ref{tomo1}) and
(\ref{tomo2}). We take the results to be the mean from the two
equations. If we do not know the symmetry of the molecule the HHG
should be measured over the whole alignment range, from
$0^{\circ}$ to $180^{\circ}$. For homonuclear diatomic molecules
the symmetry allows us to consider data from $0^{\circ}$ to
$90^{\circ}$ only. Furthermore, from the investigation of
tomographic transformations (\ref{tomo1}) and (\ref{tomo2}) using
``exact" theoretical dipole moments we found that an angular step
of $\Delta\theta=5^{\circ}$ is good enough for extracting the wave
function accurately. Thus, instead of integration over the whole
range  we use the summation over the $19$ angles from $0^{\circ}$
to $90^{\circ}$ only.

\section{Results and Discussion}

\subsection{Alignment dependence of HHG for N$_2$ and O$_2$}

In this paper, we generate HHG spectra using the Lewenstein model,
extended for molecules, as described in Sec.~II(A). In Fig.~1, we
show the HHG intensities along the direction parallel to the laser
polarization for N$_2$. Here the molecules are assumed to be fixed
in space and the directions of  laser polarization are varied from
0 to 90$^{\circ}$ (in step of 10$^{\circ}$) with respect to the
molecular axis. The laser wavelength is 800 nm. Note that our
results are for single molecules and only odd harmonics are used.
For each $(2n+1)$-th harmonic, we integrated the calculated
harmonic spectrum from the $(2n)$-th to $(2n+2)$-th harmonics. For
the tomographic imaging we need the harmonic yields in the
direction perpendicular to the laser polarization as well. Figures
2 and 3 show the alignment dependence of some selective harmonic
orders for N$_2$ and O$_2$. Both polarizations are presented. For
parallel polarization, the HHG signals are peaked near parallel
alignments for N$_2$ and about 45$^{\circ}$ for O$_2$. For
perpendicular polarization, the HHG signals of N$_2$ are largest
near 45$^{\circ}$.

To compare the HHG calculated here directly with those measured by
Itatani {\it et al.} \cite{itatani-nature} is difficult. Their data
are from the macroscopic medium such that propagation effect is
included. Furthermore, the molecules from the pump beam are only
partially aligned. Nevertheless by assuming phase matching and that
the molecules are fully aligned, they extracted the dipole moment of
the individual molecules from the macroscopic HHG. In Itatani {\it
et al.} the HHG in the perpendicular direction was not measured, but
maximal HHG intensity near 45$^{\circ}$ was used implicitly
\cite{david}. This assumption is consistent with our calculated
results for N$_2$ shown in Fig.~2.  For O$_2$, our calculations show
that the perpendicular HHG intensities almost vanish near
45$^{\circ}$ and peak at two intermediate angles near 30$^{\circ}$
and 70$^{\circ}$.

In reconstructing the HOMO's, we typically use 19 alignment angles
from 0 to 90$^{\circ}$ (in step of 5$^{\circ}$). This is also the
typical setup used in the experiments by Itatani {\it et al}
\cite{itatani-nature}. We found that using a finer angular grid do
not improve the results.

\begin{figure}
\mbox{\myscalebox{
\includegraphics{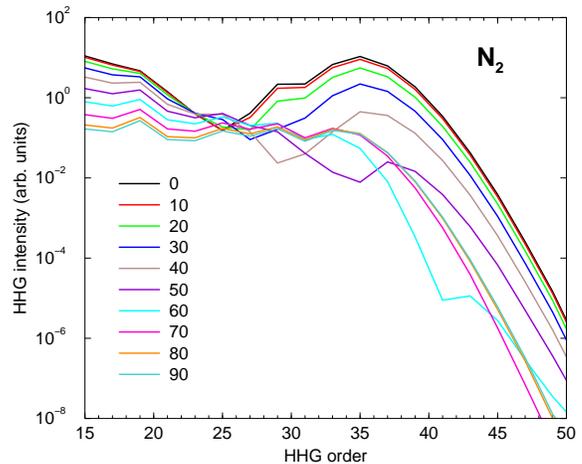}}}
\caption{(Color online) HHG spectra with parallel polarization
(with respect to the laser polarization direction) from N$_2$
aligned at 10 different angles between 0 and 90$^{\circ}$. The
laser wavelength of 800nm and intensity of $2\times 10^{14}$
W/cm$^2$ are used.} \label{fig1}
\end{figure}

\begin{figure}
\mbox{\myscalebox{
\includegraphics{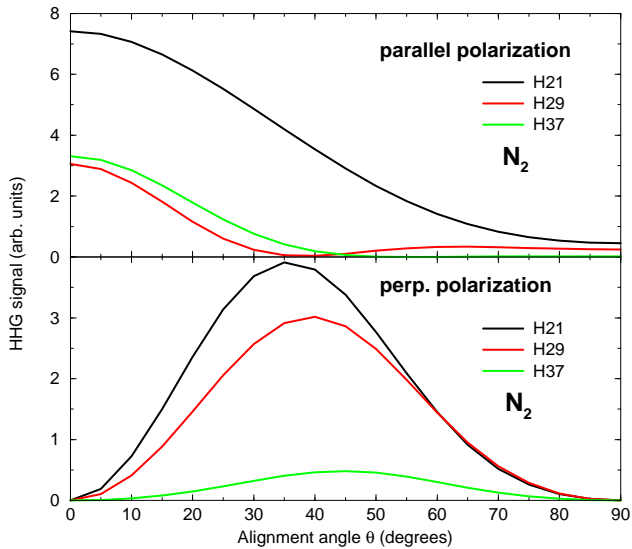}}}
\caption{(Color online) Alignment dependence of selective HHG
orders for parallel (top) and perpendicular (bottom) polarizations
from N$_2$. The laser parameters are the same as for Fig.~1.}
\label{fig2}
\end{figure}

\begin{figure}
\mbox{\myscalebox{
\includegraphics{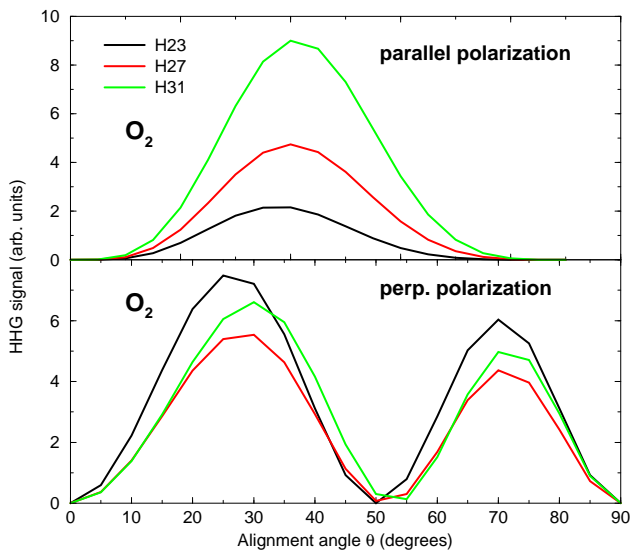}}}
\caption{(Color online) Same as in Fig.~2, but for O$_2$.}
\label{fig3}
\end{figure}

\begin{figure}
\mbox{\myscalebox{
\includegraphics{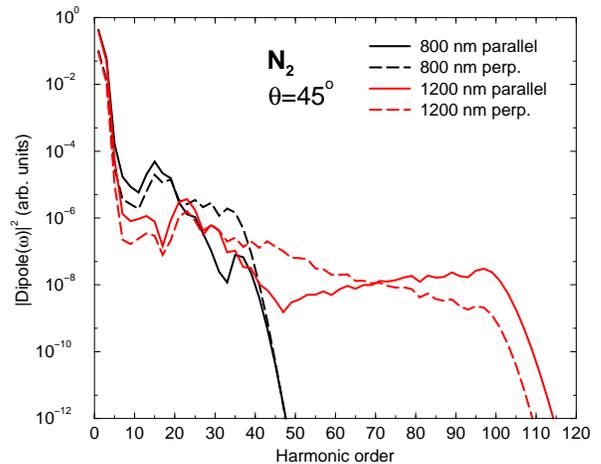}}}
\caption{(Color online) HHG spectra from N$_2$ aligned at
45$^{\circ}$ for 800 nm and 1200 nm lasers. The laser intensities
of $2\times 10^{14}$ W/cm$^2$ are used in both cases.}
\label{fig4}
\end{figure}

As we will show, in order to improve the quality of the
reconstructed HOMOs, one would like to use the HHG  over an
extended plateau region. It is better not to use  higher laser
intensities, since it would result in depletion of molecules to
distort the simple model [Eq.~(3)]. Instead, we choose to use a
laser of longer wavelength, say, of 1200 nm. We chose the same
peak intensity of $2\times 10^{14}$ W/cm$^2$, as the 800 nm laser.
In Fig.~4, we present HHG spectra from N$_2$ aligned at
45$^{\circ}$, for 800 nm (black curves) and 1200 nm (red curves).
Clearly, the plateaus are extended to much higher harmonics. We
note, however, that one harmonic order is 1.55 eV for 800 nm laser
and 1.03 eV for 1200 nm laser.

\subsection{Extracted dipole matrix elements}
\begin{figure}
\mbox{\myscalebox{
\includegraphics{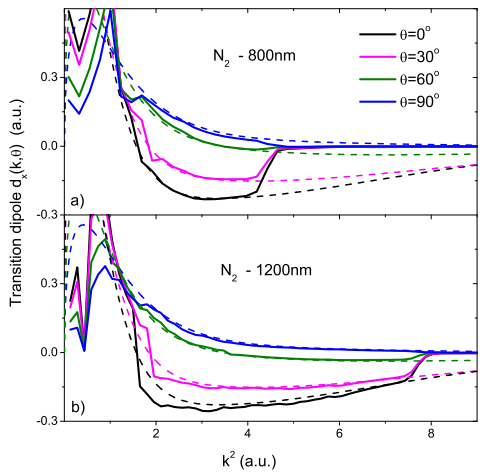}}}
\caption{(Color online) $x$-component of the transition dipoles of
$N_{2}$ obtained from HHG data (solid lines) in comparison with
the theoretical ones (dash lines) in the case of the 800 nm laser
(a) and 1200 nm laser (b).} \label{fig5}
\end{figure}

\begin{figure}
\mbox{\myscalebox{
\includegraphics{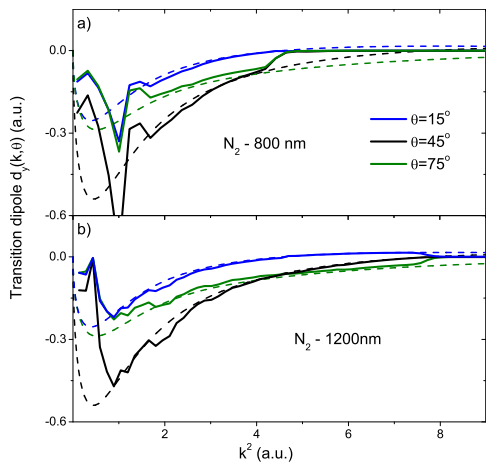}}}
\caption{(Color online) Same as in Fig.~\ref{fig5} but for
$y$-component.} \label{fig6}
\end{figure}

   Firstly we use the HHG spectra obtained in
Sec.~III(A) (to be called ``experimental" in the following) for
N$_2$ in the case of $800$ nm laser to extract the dipole moment
in the laboratory frame by using Eq.~(\ref{dipole1}). Ar(3p) (with
$I_p=15.76$ eV) is used as the reference atom, as suggested by
Itatani {\it et al} \cite{itatani-nature}. The data are available
for $19$ angles $\theta$ in the range of $[0-90^{\circ}]$. These
extracted ``experimental" dipole moments
  will be compared to those calculated theoretically. For
demonstration, in Fig.~\ref{fig5}(a) we show the ``experimental"
and theoretical results of $d_{x}(k,\theta)$ for the angles
$\theta$ of $0^{\circ}, 30^{\circ}, 60^{\circ}$ and $90^{\circ}$.
Similar results are shown in Fig.~\ref{fig6}(a) for the
$y$-component of dipole $d_{y}(k,\theta)$. In order to compare
with the theoretical data, we have normalized the $x$-component of
the dipole at $k^2=2.5$ for $\theta=0$. The normalization factor
is then fixed and used for all the other alignment angles as well
as for $y$-components. This normalization factor is also used for
N$_2$ with the 1200 nm laser.

From the Figs.~\ref{fig5}(a) and ~\ref{fig6}(a) we see that the
``experimental" dipoles compare well with the theoretical ones in
the range of $k^2=[0.9 - 4.2]$, which is equivalent to the useful
range of HHG from H17 to H47 for the $800$ nm laser. The lower
limit H17 is due to the inaccuracies in the Lewenstein model for
low harmonics, while the upper limit H47 is due to the harmonic
cut-off. For the $800$ nm laser and intensity of $2\times10^{14}$
W/cm$^2$ the cut-off position is at about H35 for N$_2$, as can
also be seen in Fig.~\ref{fig1}.

 We can enlarge the useful region of HHG by using  longer wavelength lasers. In
Figs.~\ref{fig5}(b) and \ref{fig6}(b) we show
 the ``experimental" dipole moments in the case of $1200$ nm laser with the same
 intensity. The useful dipole moments now cover the range $k^2=[1.0 - 7.5]$,
 or H27 to H113, of the $1200$ nm laser.  We note that the agreement between
 the ``experimental" and theoretical
 dipoles appears to be better for higher plateaus. This is not entirely
surprising since one should expect the simple model
Eq.~(\ref{intensity}) to agree better with Lewenstein model near the
cut-off, where the contribution from single returns dominates
\cite{rmp-review}.

\begin{figure}
\mbox{\myscalebox{
\includegraphics{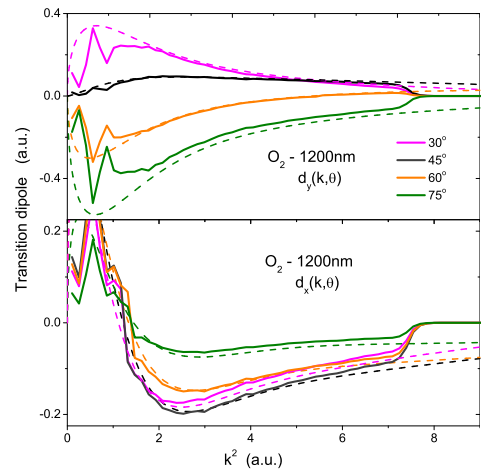}}}
\caption{(Color online) The transition dipole matrix element of
O$_2$ extracted from HHG data (solid lines) in comparison with the
theoretical ones (dash lines) in the case of the 1200 nm laser.
The upper and lower panels are for $y$ and $x$ components,
respectively.} \label{fig7}
\end{figure}

 In Fig.~\ref{fig7} we show the extracted dipole moments for
 O$_2$ molecules using the $1200$ nm laser. In this case Xe(5p)
 (with $I_p=12.13$ eV)
 is chosen as the reference atom. The normalization is
 done in the similar fashion as for N$_2$, but here we chose to normalize
the $y-$ component of the dipole at $k^2=2.0$ for
$\theta=45^{\circ}$. We see that the results are also
 comparable with the theoretical data in the range of $k^2=[1.0-7.5]$
 corresponding to the useful range of HHG from H25 to H109. The
 difference in the cut-off positions is due to the different ionization
 potentials of N$_2$ ($15.58$ eV) and O$_2$ ($12.03$ eV).

It should be noted that for O$_2$ the extracted dipole moments for
the angles near $0^{\circ}$ and $90^{\circ}$ do not agree well
with the theoretical data. It is related to the accuracy of the
small ionization rates and small HHG yields in the small angles
and near $90^{\circ}$. Note that plane-wave approximation is also
questionable near these angles, where the HOMO of O$_2$ has nodal
surfaces. However, it will be shown in the next section that these
inaccuracies do not affect much the wave function obtained by the
tomographic procedure.

\subsection{Retrieving wavefunctions by tomographic procedures}

\begin{figure}
 \mbox{\myscalebox{
\includegraphics{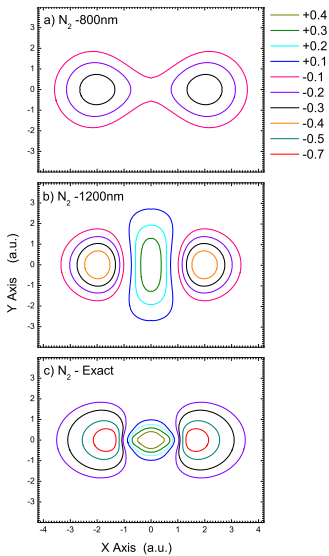}}}
 \caption{(Color online) The contour plot of the wave function of N$_2$
 obtained from HHG data in the
useful range H17 - H47 of the $800$ nm laser (a); from HHG data in
the useful range H27 - H113 of the $1200$ nm laser (b); and from
the theoretical transition dipole in the range $k^2=[0-100]$ (c).}
\label{fig8}
\end{figure}

We next examine the retrieved wavefunction from the extracted
``experimental" dipole moment. We emphasize that in the following
the wave function is understood as integrated over $z$ direction
(in the molecular frame) as given in Eq.~(\ref{integrate-wf}). In
Fig.~\ref{fig8}(a) the contour plot for the wavefunction of N$_2$
obtained from the HHG by the 800 nm laser is shown. The harmonics
used are from H17 to H47. In Fig.~\ref{fig8}(b), the wavefunction
extracted from the HHG by the 1200 nm laser is shown, using
harmonics orders from H27 to H113. The converged wavefunction
obtained from the tomographic procedure using the theoretical
dipole moments in the large range of $k^2=[0-100]$ is shown in
Fig.~\ref{fig8}(c). This is exactly the ground state molecular
wavefunction used in the calculation of the HHG. In the following
we will call the HOMOs that are obtained from the GAMESS and used
to generate HHG spectra ``exact" wavefunctions.

\begin{figure}
\mbox{\myscalebox{
\includegraphics{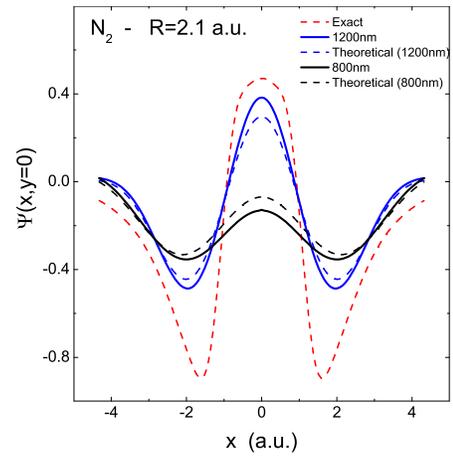}}}
\caption{(Color online) N$_2$ wave-function $\Psi(x,y=0)$. The
solid (dashed) curves correspond to the extracted (theoretical)
wave-function, respectively. The theoretical wave-function are
obtained using the tomographic procedure with the same range as
for the extracted wave-functions. The range is $k^2=[1.0-7.5]$ for
$1200$ nm laser (blue curves) and $k^2=[0.9-4.2]$ for $800$ nm
laser (black curves). For comparison we also plot the converged
wave-function obtained with the range $k^2=[0-100]$ (dashed red
curve).} \label{fig9}
\end{figure}

 By comparing the three figures in Fig.~\ref{fig8}, we see that the result
 from the 800nm is not good enough, such that even the nodal surfaces
 are not reproduced. This is due to the limited range of momentum
 space covered by the H17 to H47 harmonics used in the tomographic
 imaging. For the 1200 nm case, the retrieved wavefunction resembles
 the ``exact" one quite well, in terms of the orbital shape and the
 nodal surfaces. This clearly illustrates that the tomographic
 procedure will work better if longer wavelength lasers are used,
 such that the plateau region can be extended over many harmonics.
 For more quantitative comparison, we compare the wavefunctions along
 the internuclear axis derived by using the various approximations. In
 Fig.~\ref{fig9}, the ``exact" results are compared to the wavefunctions
 obtained  from the
 800nm and from the 1200 nm lasers. For each wavelength, we
 compare the wavefunctions derived by using the dipole moments obtained
 from the ``experimental" HHG data and from the theoretical calculations,
 but within the same range of $k^2$. Clearly the derived wavefunctions using the
 same range of $k^2$ are very close to each other. This indicates
 that the dipole moment extracted using the simple three-step model,
  Eq.~(\ref{intensity}), works well at least within the Lewenstein model.

\begin{figure}
\mbox{\myscalebox{
\includegraphics{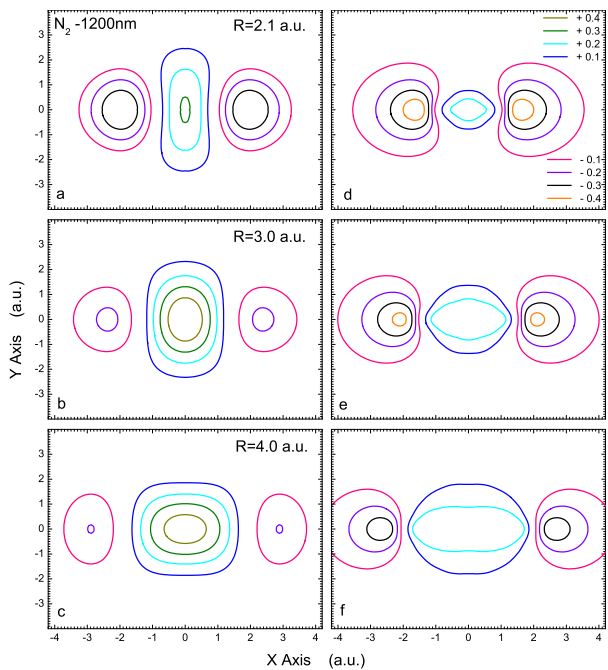}}}
\caption{(Color online) Contour plot of the wave-functions of
N$_{2}$ with the internuclear distance $R=2.1$, $3.0$, and $4.0$,
extracted from HHG data (left panels) as compared with the exact
ones (right panels).} \label{fig10}
\end{figure}

The results in Figs.~\ref{fig8} and \ref{fig9} indicate that the
tomographic imaging method works reasonably well when 1200 nm
laser is used to generate the HHG. We test the procedure for two
additional internuclear separations, $R=3$ a.u. and $R=4$ a.u. In
Fig.~\ref{fig10} we compare the wavefunctions retrieved from the
HHG, using the same range of HHG from H27 to H113. Comparing to
the ``exact" wavefunctions, we see that the positions of the nodal
surfaces, and the symmetry of the wavefunctions, are all quite
nicely reproduced. Note that in this and the following figure for
contour plot of the wavefunctions, for convenience, we have
renormalized the wavefunctions to $1$. It appears that the
tomographic procedure fails to correctly reproduce the electron
density distributions. This is especially true for $R=3$ and 4
a.u., where it overestimates the distribution between the two
nuclei. We note, however, that this is largely due to the limited
range of harmonics available. The calculations using the
theoretical dipoles with the same range of $k$ (or harmonics)
indeed show the similar trend.

\begin{figure}
\mbox{\myscalebox{
\includegraphics{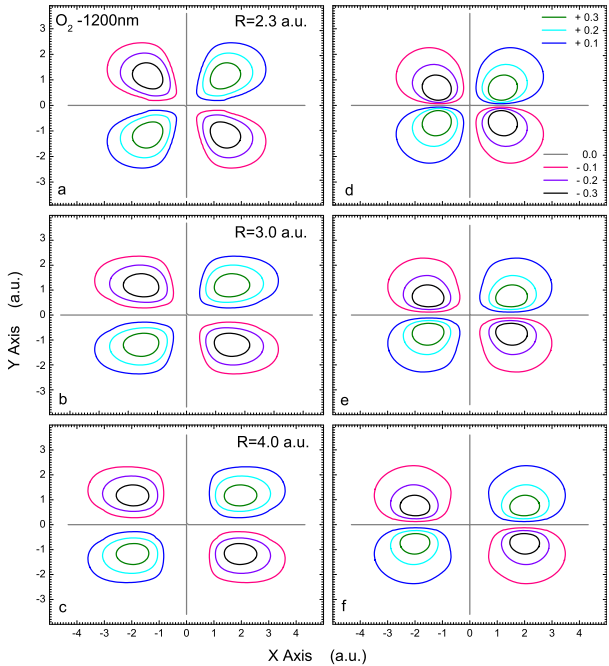}}}
\caption{(Color online) Same as for Fig.~\ref{fig10} but for O$_2$
with $R=2.3$ (top panels), $R=3.0$ (middle panels) and $R=4.0$
(bottom panels).} \label{fig11}
\end{figure}

\begin{figure}
\mbox{\myscalebox{
\includegraphics{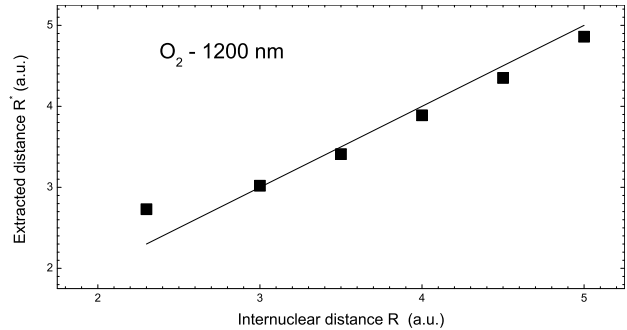}}}
\caption{Internuclear distance of O$_2$ extracted from HHG data
using the 1200 nm laser.} \label{fig12}
\end{figure}

 We further test the tomographic procedure for the O$_2$ molecule
 whose HOMO  has different symmetry ($\pi_g$ instead of $\sigma_g$
 in N$_2$ case). We used 1200 nm laser to
 generate the HHG spectra. The results for the equilibrium distance
 $R=2.3$
 and two other internuclear separations $R=3$ and 4 a.u., are shown
 in Fig.~\ref{fig11}. In this case we notice that the $\pi_g$
 orbitals are well reproduced,
 and the internuclear separations, as estimated by the distances
 between the peaks along $x$-axis, appeared to be quite accurate as
 well. In fact, we have tested the procedures for $R=3.5$, $4.5$ and
 $5.0$ a.u. as well. The deduced internuclear distances from the HHG's are
 shown in Fig.~\ref{fig12}. In Table~I, the details of the numerical results
 are shown. The input $R$, the internuclear distance $R^*$ read from the
 two peaks of the retrieved wavefunctions are compared. When the
 large range of $k^2$ (0 to 25 a.u.) in the theoretical dipoles
 is used in the tomographic
 procedure, the internuclear separations can be extracted to better
 than one percent. Using the smaller range of $k^2$ of 1.0 to 7.5
 a.u., corresponding to H25 to H109, we obtained $R^*$ accurate to a
 few percents, except for the $R=2.3$ case, which is still good
 to about 10\%.

\begin{table}[th]
\caption[]{Internuclear distance of O$_2$ extracted from HHG data
using the 1200 nm laser. $R$ - internuclear distance as input;
$R^{*}$ - extracted distance using the tomographic procedure. The
top row corresponds to the case when ``experimental" dipoles are
used, the two bottom rows - when the theoretical dipoles are
used.} \centering{ \vspace{0.2cm}
\begin{tabular}{lcccccc}
\hline \hline Input $R$ & 2.3 & 3.0 & 3.5 & 4.0 & 4.5 & 5.0 \\
\hline Extracted $R^*$, $k^{2}=[1-7.5]$ & 2.73 & 3.02 & 3.41 &3.89
& 4.35 & 4.86 \\
Theor. $R^*$, $k^{2}=[1-7.5]$ & 2.37 &
2.92 & 3.34 & 3.89 & 4.35 & 4.92\\
Theor. $R^*$, $k^{2}=[0-25]$& 2.31 &
3.01 & 3.48 & 3.98 & 4.48 & 4.97\\
\hline\hline
\end{tabular}
}
\end{table}

The relatively good agreement between the retrieved and ``exact"
wave-functions is not entirely surprising. This is because
Eq.~(\ref{intensity}) can be considered as an approximation to the
Lewenstein model used to generate our ``experimental" data. In
particular, this agreement clearly cannot be used to justify the
plane-wave approximation, since this approximation is used in both
models. On the other hand, one should keep in mind that the
Lewenstein model has been shown to be able to interpret the main
features of the measured HHG spectra for atoms and simple
molecules \cite{rmp-review,zhou,atle06}. We will discuss in
details the limitations of this approach in the next section.

\section{Intrinsic and practical limitations of the tomographic
procedure} In this section we reexamine the assumptions made in
the tomographic procedure, following  Itatani {\it et al}
\cite{itatani-nature}. The starting point of the method is  the
Fourier slice theorem, Eqs.~(6) and (7), which gives a
mathematical identity between $\Psi(x,y)$ and the dipole matrix
elements ${\bm d}({\bm k})$ in the direction parallel and
perpendicular to the laser's polarization, with the condition that
the transition dipole ${\bm d}({\bm k})$ be given by $\langle{\bm
k}|{\bm r}|\Psi_0({\bm r})\rangle$, i.e., the final continuum
states be represented by plane waves. Operationally, the dipole
matrix elements are to be obtained from experiments, for example,
using high-order harmonics generated from aligned molecules, with
laser's polarization direction varying from parallel to
perpendicular, with respect to the alignment direction of the
molecules, as carried out by Itatani {\it et al.}

 We now examine the assumptions and limitations encountered in the
 practical implementation of the tomographic imaging method.

 (1) Limitation of the available experimental spectral range.

 The Fourier slice theorem requires that the dipole matrix
 elements be measured for the whole spectral range, while the
 usable range from the HHG is limited to the small plateau region.
 In Figs.~8-11 we show the effect on the retrieved wavefunctions if truncated
 spectral range is used. For the small plateau region available
 from HHG generated by 800 nm lasers, we conclude that accurate
 wavefunctions cannot be obtained, but
 significant improvement can
 be made if the HHG generated by 1200 nm or longer wavelengths are
 used. We comment that in general by introducing a filter
 function, the accuracy of the retrieved HOMO can be
 improved \cite{kak,patch06}.

 (2) Limitation of the approximate three-step model for extracting molecular transition
 dipoles from the measured HHG spectra.

 The link between the HHG yields and the
 ``experimental" dipole moments of an individual molecule which are
  used in retrieving the molecular wavefunction of the
 initial state is given by Eq.~(3) -- an equation based on the three-step model
 of the high-order harmonic generation. Eq.~(3) was not derived
 from a rigorous HHG theory so its validity can only be checked
 empirically. In Eq.~(3), propagation effect was included by
 assuming perfect phase matching, an approximation which will be addressed later.
 Given that Eq.~(3) is correct, how to interpret the derived
 ``experimental" dipole matrix element $|{\bm d}(\omega,\theta)|$? In Itatani {\it et al.},
 they interpreted it as
the dipole matrix element to be used in the Fourier slice theorem,
Eqs.~(6) and (7). This amounts to approximating the ``exact"
continuum states of a molecule by plane waves over the whole
spectral range, a very poor approximation well known from
photochemistry.

In this paper, we showed that Eq.~(3) is reasonably valid for HHG
generated from a single molecule if the harmonics are calculated
using the Lewenstein model, see the comparisons in Figs.~5-7. In
the Lewenstein model, the continuum wavefunction is approximated
by plane waves, thus the ``experimental" dipole matrix element
from Eq.~(3) can be used directly in Eqs.~(6) and (7) to retrieve
the molecular wavefunction. Does Eq.~(3) permit us to extract
accurate dipole matrix elements? We do not know yet. Even if it
does, it would not help since the tomographic procedure employs
the condition that the continuum states be represented by plane
waves.

A further limit  of Eq.~(3) is that it gives only the absolute
values of the dipole matrix element. For simple linear molecules,
the dipole matrix elements are real such that one can just change
the sign whenever the absolute value goes through a near-zero
minimum, as adopted by Itatani {\it et al}. For more complex
molecules, the dipole moments can become complex and it would be
difficult to extract the phase information. As for measurement of
the relative phase of high harmonics, important progress has been
achieved recently \cite{agostini,mairesse,salieres}.

(3) Limitations from the macroscopic effects.

  The high-order harmonics measured in the laboratory are from the
  coherent emission of light from many molecules. Eq.~(3) was
  written down under the perfect phase matching condition, but
  how well is it  met in actual experimental condition, especially from
  partially aligned molecules? We note
  that in applying Eqs.~(6) and (7), both parallel and perpendicular
  components of the HHG are needed. This can be done in principle if the molecules are
  fully aligned, i.e., all are pointing in a fixed direction in
  space. Such condition cannot be reached by the pump laser pulse
  where molecules are only partially aligned. Thus the angles between
  the laser's polarization and the  molecules in the ensemble are
  not all
  the same, making it very difficult to extract the perpendicular component of the
  dipole moment of the individual molecule from the measured HHG signal. Note
  that for a unaligned ensemble of molecules, the perpendicular
  component of HHG vanishes.
  In Itatani {\it et al.}, the molecules were assumed to be  fully
  aligned in the direction of the pump pulse. They did not measure
  the perpendicular component of HHG, but the perpendicular component
  of the dipole moment was assumed. It turns
  out that the form they assumed is very close
  to what we calculated using the Lewenstein model, shown in Fig.~2(b).
  Note that in our present model study we did not face the
    macroscopic issues since we derived ``experimental" dipole
    matrix elements from HHG due to a single molecule fixed in
    space only.

\section{Chemical imaging with high-order
harmonic generation}

\subsection{Alternative methods of chemical imaging}

 In Section III.C, with our most optimistic assumptions, we
 have shown that  the HOMO wavefunction cannot be accurately
 retrieved based on the tomographic method since the spectral
 range provided by the plateau harmonics is too limited. On the other hand, the
 symmetry of the HOMO, and the internuclear distances between the two atoms,
 are accurately reproduced from the retrieved wavefunction, especially
 when lasers with longer wavelengths, say, 1200 nm,  are
 used. Such information is  comparable to    what one can obtain from
 electron or X-ray diffraction experiments. On the other hand,
 since HHG can be generated with
femtosecond lasers of duration of the order of 10 fs or less, the
retrieval of even such limited information can be very useful in
probing the time dependence of a chemical transformation at time
resolution of the order of tens to sub-ten femtoseconds. Thus
further investigations of possible dynamic chemical imaging with
HHG by femtosecond lasers are warranted.

 In the future, however, we believe that dynamic chemical imaging
 be developed along a different route. As discussed in Section IV,
 the steps involved in obtaining single molecule dipole moments
 from the macroscopic HHG  rely on many assumptions which
 are difficult to remove. At present, we suggest that dynamic
 chemical imaging with HHG  be proceeded as follows. We assume that
 the initial configuration of the molecule in the ground state is
 known from the conventional imaging method. Experimentally
 the alignment dependent HHG from the ground state of such a
 molecule will first be measured, and theoretical calculations be
 performed to make sure that the
 experimental data at this initial configuration are well described.
 Under chemical transformation, our goal is to locate the
 positions of the atoms as they change in time, by measuring their HHG
spectra at different time delays, following the initiation of the
reaction. Assuming that such HHG data are available
experimentally, the task of dynamic chemical imaging is to find
ways to extract the intermediate positions of all the atoms during
the transformation, and in particular, to identify the important
 transition states of the reaction. This will be done by the
 iterative method.

 In the iterative method, we will first make a guess on the new
 positions of all the atoms in the molecule. A good guess is
 to follow the reaction coordinates along the path where the
 potential surfaces have the local minimum. Existing quantum
 chemistry codes (see, for example, Ref.~\cite{gamess,gaussian})
 provide good guidance as a starting point. For each initial guess
 the HHG from single molecules will be calculated. Then the
 propagation effect and the alignment of molecules are taken into
 account. The resulting macroscopic HHG are then compared to
 the experimental data to find atomic configurations that
  best fit the experimental data. Recall that for each time
 delay, the harmonics can be measured for many different angles
 between the aligning laser and the probe laser. Different
 wavelengths of the lasers can also be used. At present, we
 believe that HHG be measured   along the probe laser's
 polarization direction only, since the perpendicular component
 tends to be severely averaged out when molecules are only
 partially aligned. In principle, the phase of each harmonic can
 also be measured experimentally \cite{salieres}. For the chemical imaging
 purpose, however, we think it is easier to use different wavelengths to
 generate the larger data set for retrieving the atomic
 configurations instead. We mention that the best fitting criterion
 of the iterative method is still to be developed, but at least one
 can start with the genetic algorithm \cite{genetic}.

  For the methods suggested above to work, a relatively accurate and efficient
  theoretical model for calculating HHG from single molecules is
  needed. Direct solution of the time-dependent Schr{\"o}dinger equation
  for molecules in a laser field, even within the single active
  electron model is out of questions. On the other hand, a model
  such as the Lewenstein model is very efficient. It is
  straightforward to include the propagation effect and the
  anisotropic distributions of molecules. Thus if the macroscopic
  HHG's can be calculated efficiently with reasonable accuracy, the large
  set of experimental HHG data can then be analyzed to retrieve the
  atomic configurations at each time step efficiently. We emphasize
  that in this iterative procedure, the number of parameters that
  are to be determined is small since only the positions of the
  atoms that undergo large changes are to be determined.

  We summarize that with this method, there are a number of advantages over the
  tomographic method: (1) No need to assume that the continuum states
  be (incorrectly) represented by plane waves; (2) No obvious need
  to measure HHG along both the parallel and the perpendicular
  directions. (3) The method is based on a fitting procedure, not a
  Fourier transformation, such that limited spectral range like the plateau harmonics may be
  adequate. For the tomographic method the whole spectral range is needed.
  (4) The propagation and the alignment effect of the
  molecules can straightforwardly be incorporated in the fitting process.

  \subsection{Outlook of chemical imaging with HHG}

  The method we have suggested above for chemical imaging formally
  requires an accurate and efficient theoretical method for
  calculating the HHG of a complex molecule. In a chemical reaction
  involving ``complex" molecules, normally  only  a few
  atoms  make large change in positions. Such changes often modify the
  HOMO's  significantly  since the symmetry of the molecule is altered,
  resulting in significant change in the alignment dependence of the HHG. We note that
  such changes are typically accounted for by the Lewenstein model
  already. Thus even before a more accurate theory of HHG becomes available, the
  procedure outlined in the previous subsection can be tested at
  least qualitatively.

  As an example, consider the isomerization of acetylene to
  vinylidene \cite{c2h2ref0,c2h2ref1,c2h2ref2}. We have calculated the simplified
  possible reaction path of the
  isomerization  by using  density functional theory~\cite{dft} with
   the hybrid exchange-correlation functional of Perdew, Burke and
Ernzerhof (PBE1PBE)~\cite{pbe1pbe} and Dunning's correlation
consistent basis set AUG-cc-pVTZ~\cite{gaussian}. It is in excellent
agreement with Ref.~\cite{c2h2ref2}.  This path involves the
well-known transition state TS2~\cite{c2h2ref1}.
  To reach TS2 from acetylene, at least about 2 eV is needed.
  Before reaching the vinylidene there is one additional local
  minimum (LM2) and one additional transition state (TS1) in the potential
  surface. The positions of the atoms and the shape of the HOMO for each of
  these special configurations are shown in the figure. Note that
  the binding energy and the HOMO for acetylene is very different
  from the other four. Such differences can be easily revealed from
  the ``alignment" dependence of HHG. In Fig.~13 we show the HHG
  calculated for these different isomers. The alignment angle is
  defined to be between the C-C axis and the laser polarization
  direction. Except for the linear acetylene, the HHG for the other
  planar molecules should depend on the
 $\phi$ angle which is   between the plane of the molecule
and the plane  of the C-C axis and the laser's polarization. In
Fig.~13, this $\phi$-dependence has been integrated over.


\begin{figure}
\begin{center}
\mbox{\myscalebox{
\includegraphics{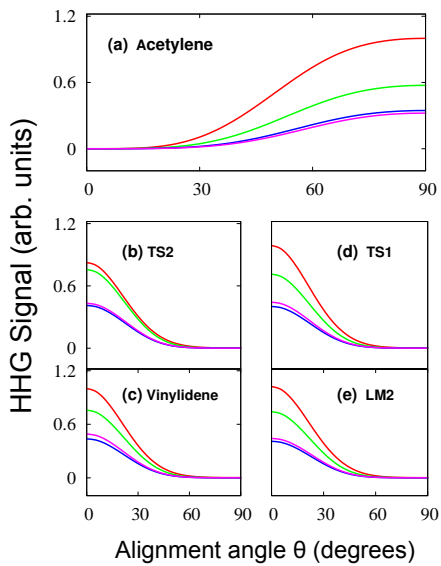}}}
\caption{(Color online) Predicted angular dependence of the 21th
(red), 25th (green), 29th (blue), 31st (purple) harmonic yields in
the plateau region for C$_{2}$H$_{2}$ isomers: (a) Acetylene; (b)
TS2; (c) Vinylidene; (d) TS1; and (e) LM2.} \label{fig15}
\end{center}
\end{figure}


\begin{figure}
\mbox{\myscalebox{
\includegraphics{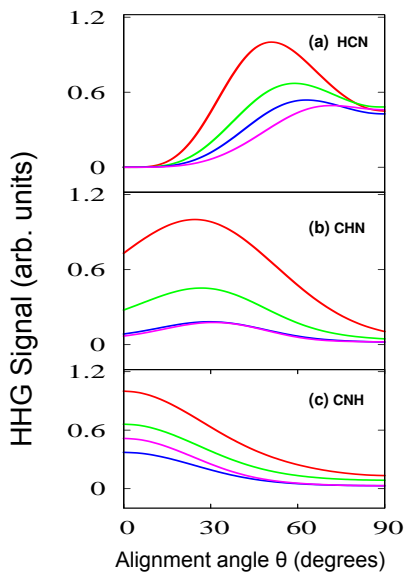}}}
\caption{(Color online) Predicted angular dependence of the 21th
(red), 25th (green), 29th (blue), 31st (purple) harmonic yields in
the plateau region for HCN isomers: (a) hydrogen cyanide HCN; (b)
transition state CHN; and (c) hydrogen isocyanide CNH.}
\label{fig17}
\end{figure}

   The results in Fig.~13 illustrates the potential and the
   limitations of using the HHG for imaging molecules. While it is
   relatively easy to identify acetylene from vinylidene by the
   alignment dependence of the HHG yields, it is difficult to
   distinguish vinylidene from  other transient molecules, TS1, TS2 and LM2.
   Although the atomic arrangements for
   TS2, LM2 and TS1 are quite different from vinylidene,
   major features of the HOMO from each of these ``molecules" are relatively
   similar. Since the HHG, at least within the simplest model, is
   generated from the HOMO only, it is to be understood as a probe
   of the highest occupied molecular orbital, rather than a probe of the
   positions of the atoms in the molecule, as from the diffraction method.
   Thus, in a way, imaging by HHG is more similar
   to (e,2e) momentum spectroscopy~\cite{weigold,eland}. While it may be
   possible to use HHG to distinguish vinylidene from TS2, LM2 and TS1,
   by carefully analyzing the
   dependence of HHG on the alignment angle $\theta$ and the
   azimuthal angle $\phi$, or by using lasers with different wavelengths,
   the procedure is likely much
   more demanding. On the other hand, in a typical reaction,
     a vibrational wave packet is created. Since the
   potential surfaces of these four states are quite close to each
   other, unique identification of the atomic configurations of these
   four isomers as
   stationary states may be an oversimplification. Nevertheless, other
   imaging methods would be needed to map out the full evolution
   from acetylene to vinylidene.  The transformation away from the
   acetylene, on the other hand, would be very easy to identify
   using the HHG.

   Before closing, we offer one more example -- the isomerization
    of hydrogen cyanide HCN to hydrogen isocyanide HNC \cite{hcnref0}.
    We have calculated the local potential minima, the shape of the HOMO,
    the atomic configurations, and the contour plot of the potential
    energy surface calculated by using DFT~\cite{dft} with Becke's
   three parameter hybrid functional B3LYP~\cite{b3lyp} and
   6-311+G(2df,2pd) Gaussian-type basis set~\cite{gaussian}.
   The latter provides information on the reaction path from the linear HCN
   to the linear HNC,  via a transition state of a coplanar molecule
   as sketched. A fair large energy of 2.1 eV is needed to reach the
   transition state. Note that the HOMO's for the three ``molecules"
   are rather different, and these differences are easily revealed from
   the large difference in the alignment dependence of the HHG yields
   shown in Fig.~14, calculated again using the Lewenstein model.
   Imaging the transition from HCN to HNC via the transition state
   using the HHG would appear to be much simpler in this case.
   (We comment that in this example, Coulomb explosion technique probably
   would be the easiest since there are only three atoms in the
   system.)

   In the discussion so far, the molecules have been assumed to be
   stationary with its major axis making an angle $\theta$ with
   respect to the laser polarization. Linear molecules can be aligned
   by a short linear-polarized laser, or oriented in a combined DC
   field and a laser \cite{sakaiprl}. For more complex molecules, they can be
   oriented with a elliptically polarized laser pulse \cite{stapelfeldt}. Clearly the
   orientation and alignment of the molecules itself is a
   challenging subject on its own. With the rapid progress of
   laser technology, we are optimistic that HHG from aligned
   and/or oriented molecules will become available for dynamic
   imaging in the coming years.

\section{Summary and Future Directions}

  In this paper we examined the theoretical basis of the tomographic method used
  by Itatani {\it et al.}~\cite{itatani-nature}.   Due to the limited spectral
  range provided
  by the high order harmonics in the plateau region, we show that
  intrinsically it is very difficult to reconstruct accurately the
  wavefunction of the highest
  occupied molecular orbital (HOMO) from the measured
  high order harmonics, but the symmetry of the HOMO and
  the internuclear separation  can indeed be adequately obtained.
  The tomographic method relies on extracting the dipole
  moment of   individual molecules from the HHG measured in a macroscopic medium,
  and that the continuum
  wavefunctions in the dipole moment of a molecule be represented by plane
  waves. The former posts many practical difficulties owing to the
  macroscopic propagation effect and the imperfect alignment of
  molecules in the laser field. For the latter, representing
  continuum wavefunctions of a molecule by plane waves is a
  drastic
  simplification. These limitations pose severe obstacles in any attempt to
  improve the tomographic imaging of molecules.

   In spite of these limitations, we showed that useful
   information on the molecule, in particular, the nature of the
   HOMO and   the position  of the atoms in a transient molecule,
     can be extracted from the measured HHG's. In view of the
     limitations of the tomographic method, we have
   proposed an iterative approach to extract such information in
   the future. The method relies on an   efficient and relatively
   accurate theoretical model for
   calculating HHG from molecules. With the availability of such a
   theory, the propagation effect and other macroscopic effects
   can be modelled to generate HHG that can be compared to the measured data.
   However, such a simple theory for the HHG by molecules is not yet
   available today. On the other hand, the alignment dependence of
   the HHG emission is very sensitive to the nature of the HOMO.
   If the HOMO changes significantly when the molecule  undergoes
   transformation, for example, in isomerization, then the HHG yield
   would   reveal  the changes efficiently. Using HHG calculated
   from the Lewenstein model for the isomerization of  C$_2$H$_2$ and of
   HCN, we showed that dynamic chemical imaging with HHG is likely
   possible even without detailed quantitative iterative
   procedures outlined in Section V.

   Still many challenges remain  for dynamic chemical imaging with lasers.
   Theoretically it
   is imperative that a simple and accurate theory of HHG from
   molecules be developed. Experimentally, techniques for
   efficiently   aligning molecules with lasers are needed. Since the HHG yield
   from each molecule is small, the fraction of molecules that
   undergo chemical transformation should be large within the
   target volume. Because HHG can also be generated from the
   unexcited ground-state molecules, the intensity of the probe
   lasers should be selected appropriately.  One advantage of
   using HHG for chemical imaging is that it actually probes the
   HOMO instead of the atomic positions of the whole molecule,
   thus it is much more directly relevant to chemical transformation
   since it is the change of the outermost electron orbital that dictates
   chemical reactions.

\section*{Acknowledgement}

We acknowledge the discussions with David M. Villeneuve and Paul
B. Corkum on the details of their tomographic procedure as well as
their results \cite{patch06,david06} prior to publication. This
work was supported in part by Chemical Sciences, Geosciences and
Biosciences Division, Office of Basic Energy Sciences, Office of
Science, US Department of Energy.

\end{document}